\documentclass[12pt,reqno]{article}
\usepackage{url}

\usepackage{comment}

\usepackage{graphicx,float}

\usepackage[style=apa, citestyle=authoryear,maxbibnames=99, giveninits=true, maxcitenames=2, mincitenames=1, uniquelist=false, uniquename=init, url=true, doi=true]{biblatex}

\renewbibmacro{in:}{} 
\DefineBibliographyStrings{english}{%
    andothers = {\em et\addabbrvspace al\adddot}
}

\usepackage{amsmath,amsfonts,amssymb,mathtools}

\usepackage[bookmarks=false]{hyperref}

\usepackage{eurosym}

\newcommand{\TF}{\mathrm{TF}}
\newcommand{\IDF}{\mathrm{IDF}}
\newcommand{\TFIDF}{\mathrm{TF \text{-} IDF}}
\newcommand{\CS}{\mathrm{CS}}

\usepackage{geometry}
\usepackage{styles/orcidlink}

\begin{document}


%
\title{A text analysis for\\
 Operational Risk loss descriptions\footnote{
 \scriptsize{
 This research is a part of the project entitled ``The state of the art in anomaly detection and model construction, with the focus on natural language processing (NLP) in actuarial modelling'' which has been supported by the Committee on Knowledge Extension Research (CKER) of the Society of Actuaries (SOA) Research Institute, and the Casualty Actuarial Society (CAS).
 }
 }}
 
%
%

\author{
  \small{Davide Di Vincenzo} \orcidlink{0000-0002-3413-5515}\\
  \scriptsize{Group Non-Financial Risks, UniCredit S.p.A., Milano, Italy}\\
  \scriptsize{\texttt{Davide.DiVincenzo@unicredit.eu}}
  \and
  \small{Francesca Greselin} \orcidlink{0000-0003-2929-1748}\\
  \scriptsize{Department of Statistics and Quantitative Methods}\\ 
  \scriptsize{University of Milano Bicocca, Milano, Italy}\\
  \scriptsize{\texttt{francesca.greselin@unimib.it}}
  \and
  \small{Fabio Piacenza} \orcidlink{0000-0003-4278-3624} \\
  \scriptsize{Group Non-Financial Risks, UniCredit S.p.A., Milano, Italy}\\
  \scriptsize{\texttt{Fabio.Piacenza@unicredit.eu}}\\ 
  \scriptsize{University of Milano Bicocca, Milano, Italy}\\
  \scriptsize{\texttt{f.piacenza1@campus.unimib.it}}
  \and
  \small{Ri\v{c}ardas Zitikis} \orcidlink{0000-0002-2663-0030}\\
  \scriptsize{School of Mathematical and Statistical Sciences}\\
  \scriptsize{Western University, London, Ontario}\\
  \scriptsize{\texttt{rzitikis@uwo.ca}}
}

\maketitle              

\vspace{-5mm}

\begin{abstract}

Financial institutions manage operational risk (OpRisk) by carrying out activities required by regulation, such as collecting loss data, calculating capital requirements, and reporting. 
For this purpose, for each OpRisk event, loss amounts, dates, organizational units involved, event types, and descriptions are recorded in the OpRisk databases. 
In recent years, operational risk functions have been required to go beyond their regulatory tasks to proactively manage operational risk, preventing or mitigating its impact. As OpRisk databases also contain event descriptions, an area of opportunity is to extract information from such texts. 
The present work introduces for the first time a structured workflow for the application of text analysis techniques (one of the main Natural Language Processing tasks) to the OpRisk event descriptions to identify managerial clusters (more granular than regulatory categories) representing the root-causes of the underlying risks. 
We have complemented and enriched the established framework of statistical methods based on quantitative data. 
Specifically, after delicate tasks like data cleaning, text vectorization, and semantic adjustment, we have applied methods of dimensionality reduction and several clustering models with algorithms to compare their performances and weaknesses. 
Our results improve retrospective knowledge of loss events and enable to mitigate future risks. 

\end{abstract}

{\small
\textbf{Keywords:}
}
{\footnotesize
clustering, natural language processing, operational risk, text analysis  
}

\section{Introduction}

The operational risk (or OpRisk) is defined as the risk of loss resulting from inadequate or failed internal processes, people and systems, or from external events, and also includes the legal risk (\citeauthor{CRR}, \citeyear{CRR}).
International financial institutions typically manage this risk inside specific operational risk management functions, which perform the activities prescribed by the regulations, such as:
\begin{itemize}
\item Data collection (\textit{e.g.}, loss data, scenario analysis and risk indicators)
\item Capital requirement calculations using Advanced Measurement Approach (AMA) internal models
\item Reporting of loss data
\end{itemize}
In order to perform the above mentioned activities, financial institutions have to define and implement databases to collect and store the necessary information.
In the case of loss events due to operational risk, at least the following attributes are collected:
\begin{itemize}
\item Loss amounts
\item Dates (occurrence, discovery and accounting)
\item Affected organizational units
\item Basel loss event types (Internal Fraud; External Fraud; Employment Practices and Workplace Safety; Clients, Products \& Business Practices; Damage to Physical Assets; Business Disruption and System Failures; Execution, Delivery \& Process Management)
\item Event descriptions
\end{itemize}
The operational risk databases contain the above mentioned structured data, which are used for regulatory activities. 
However, during the last years, the operational risk functions have been increasingly required to move beyond their regulatory tasks, providing a more effective contribution in order to pro-actively manage the risk, and prevent or mitigate its impact.
In particular, the databases contain the operational risk event descriptions, which are usually defined as free text fields. 
The possibility to make all the information available to the operational risk analysts, including those from the loss event descriptions, represents great opportunities to improve the knowledge about loss events and to also design the most adequate mitigation strategies.

The present work is among the first ones that have addressed the application of text analysis techniques to the OpRisk event descriptions.
Text analysis, together with speech recognition and automatic translation, is one of the main tasks of the Natural Language Processing (NLP), which is a branch of the Artificial Intelligence (AI). 
In particular, to the best of our knowledge, for the first time in literature, the present paper defines a general structured workflow that can be applied to the operational risk descriptions to analyze them for several purposes.

\begin{figure}
    \centering
    \includegraphics[width=\textwidth]{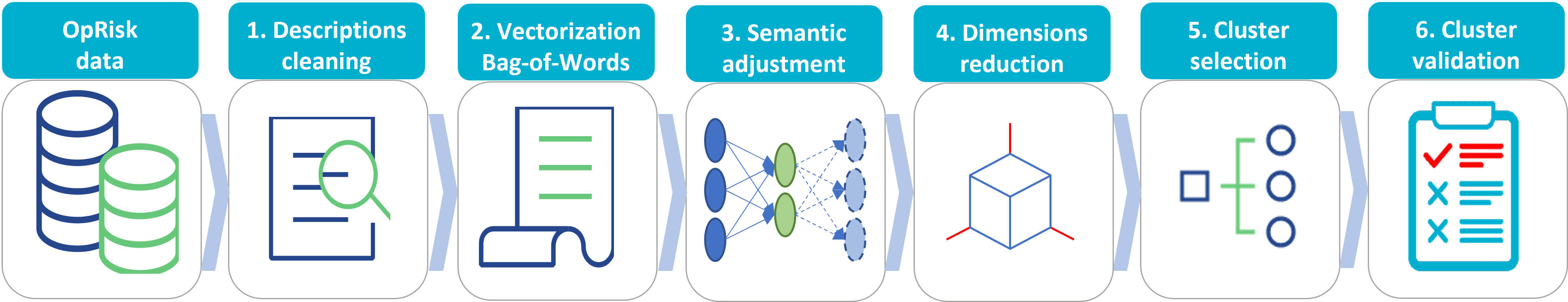}
    \caption{Workflow for operational risk descriptions analysis.}
    \label{WorkflowChart}
\end{figure}

The proposed workflow includes the following steps (as represented in Figure \ref{WorkflowChart}):
\begin{enumerate}
    \item Description cleaning (\textit{e.g.}, splitting of different languages,
    removing stop-words, reducing words to their lemmas).
    \item Text vectorization (building a document-by-term matrix, where each element is properly weighed).
    \item Semantic adjustment (enriching the document-by-term matrix, considering the semantic similarity among words).
    \item Dimensionality reduction (building a 2D representation of the data, where each point is an event description, and similar ones are represented as clusters of points).
    \item Cluster selection (points within each cluster can be tagged by the OpRisk analysts). 
    \item Cluster validation (application of clustering and classification techniques to validate and support the clustering performed by the analysts).
\end{enumerate}
For the first time, the current work addresses the applications of text analysis 
to the OpRisk loss event descriptions, proposes a structured workflow in order to define an overarching framework, thus complementing the one already applied on quantitative data.

Regarding the following parts of this work, Section \ref{Literature} gives a literature review of the text analysis applied to operational risk.
Section \ref{Workflow} describes in detail the steps of the proposed workflow.
Section \ref{Application} reports an application of the proposed workflow to the descriptions of the CoRep Operational Risk data set for UniCredit banking group, where all the data elaborations and analyses have been performed using software R (\citeauthor{R}, \citeyear{R}). 
Finally, Section \ref{Conclusion} summarizes the main achievements and results of this work, discussing also possible extensions in several directions.

\section{Literature review}
\label{Literature}

For the framework and analyses applied on quantitative operational risk data, including related challenges, one can refer to, for example, \citeauthor{soprano2010measuring} (\citeyear{soprano2010measuring}),
\citeauthor{CopeMignola} (\citeyear{CopeMignola}),
\citeauthor{Shevchenko} (\citeyear{Shevchenko}),
\citeauthor{shevchenko2006bayes} (\citeyear{shevchenko2006bayes})
\citeauthor{danesi2016} (\citeyear{danesi2016}), 
and \citeauthor{bazzarello2006} (\citeyear{bazzarello2006}).

Moving to qualitative data, the existing literature proposes only a few solutions for the analysis of textual data related to OpRisk loss event descriptions.

\citeauthor{Pakhchanyan_2022} (\citeyear{Pakhchanyan_2022}) apply machine learning techniques to operational risk descriptions, in order to automatically classify events into Basel event types.
It is worth noting that this article adopts supervised methods to classify OpRisk events into pre-defined regulatory categories, but it does not propose solutions to identify new managerial (more granular) clusters that can be used to understand the root-causes of the underlying risks.
The classification of OpRisk events is also discussed by \citeauthor{ZHOU20211} (\citeyear{ZHOU20211}), who propose semi-supervised methods to include unlabeled data in the training stage.

\citeauthor{WANG2018136} (\citeyear{WANG2018136}) and \citeauthor{Wang_2022} (\citeyear{Wang_2022}) investigate the main operational risk factors, applying the Latent Dirichlet Allocation (LDA), but without reporting many details on the applied descriptions cleaning and text vectorization.

\citeauthor{Sadeghi_2019} (\citeyear{Sadeghi_2019}) provide a preliminary proof-of-concept for the potential usefulness of statistical and NLP approaches in operational risk modelling, applying LDA and long short-term memory neural networks (LSTM).

\citeauthor{ORX_2019} (\citeyear{ORX_2019}) suggest that text analysis methodologies can be useful to gain deeper insights into the operational risk data, although without proposing detailed approaches.

A recent literature review on the application of text analysis in the financial sector (\citeauthor{su11051277}, \citeyear{su11051277}) reveals that the main research focus is on stocks price prediction, financial fraud detection and market forecast.
In the literature, there are several proposals to manage fraud risk (which is a part of OpRisk) by making use of text analysis.
For example, \citeauthor{Holton_2009} (\citeyear{Holton_2009}) proposes a methodology to detect financial frauds, identifying and classifying emails with disgruntled communications.

\section{Workflow for OpRisk descriptions analysis}
\label{Workflow}

\subsection{Descriptions cleaning}
\label{Cleaning}

Descriptions have to be prepared for the analysis using some cleaning procedures.
The set of all descriptions (or documents) to be analyzed is called “corpus”. 
Procedures to clean texts include the following ones:
\begin{itemize}
    \item Data anonymization: applying routines to retrieve and delete (or substitute with conventional tags) any personal information and dates from texts, for compliance with GDPR (\citeauthor{GDPR}, \citeyear{GDPR}) and for analytical purposes (\citeauthor{francopoulo:hal-02939437}, \citeyear{francopoulo:hal-02939437}).
    \item Splitting of different languages: applying routines to recognize and separate parts of text written in different languages (\citeauthor{Jauhiainen_2018}, \citeyear{Jauhiainen_2018}).
    \item Ignoring cases, which can be done by case-folding each letter into lowercase. 
    \item Removing punctuations and digits.
    \item Removing frequent words that do not contain much information, called stop-words, like articles, pronouns, conjunctions, and words like “of”, “about", “that", etc. 
    Lists of stop-words are readily available for the main languages.
    In particular, for the application described in Section \ref{Application}, the stop-word list, related to the English language, has been derived from this source: \url{https://metacpan.org/pod/Lingua::StopWords}.
    \item Using regular expressions to detect special characters (\textit{e.g.}, “ù”, “ä”, “$|$”, etc.) and remove them.
    \item Reducing words to their lemmas (\textit{e.g.}, “pay” from “paying”, “client” from “clients”), substituting each word with the corresponding canonical form. 
    The lemmatization reduces the number of distinct words in a text corpus and increases the frequency of occurrence for some of them.
\end{itemize}

\subsection{Text vectorization - Bag-of-Words (BoW)}

\begin{figure}
    \centering
    \includegraphics[width=\textwidth]{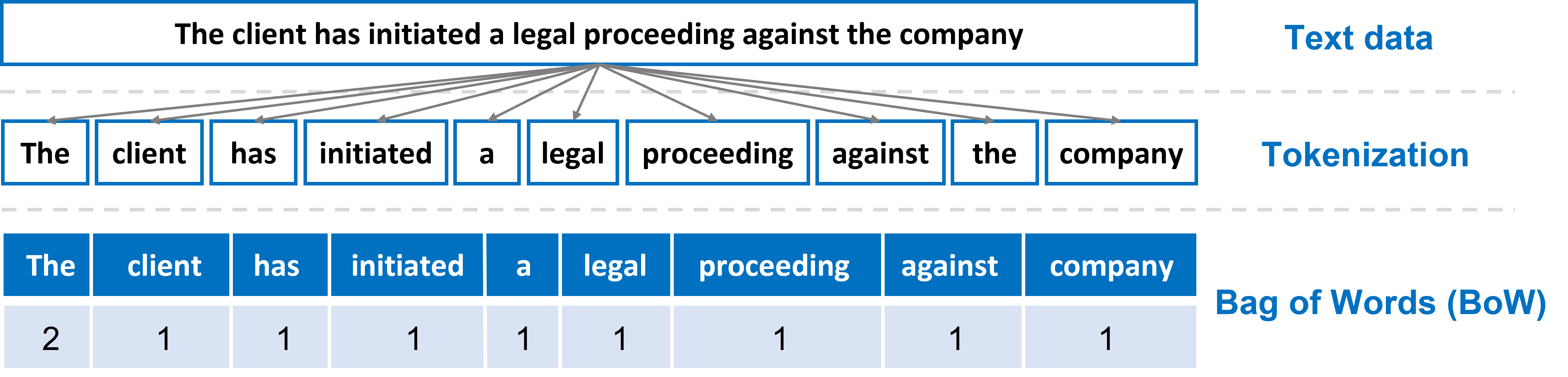}
    \caption{Example of Bag-of-Words representation (without stop-words removal).}
    \label{Example_BoW}
\end{figure}

According to the BoW approach (\citeauthor{harris54}, \citeyear{harris54}), the data set is transformed into a matrix, where:
\begin{itemize}
    \item The row $i$ represents the $i$-th document $d_i$
    \item The column $j$ represents the $j$-th term $w_j$ of the transformed data set
    \item In the cell $(i,j)$ of the document-by-term matrix, we store the Term Frequency (TF) $\TF(d_i,w_j)$ of the term $w_j$ in the document $d_i$ (\textit{i.e.}, the number of times $w_j$ appears in $d_i$) (\citeauthor{5392697}, \citeyear{5392697})
\end{itemize}
An example of BoW representation is given in Figure \ref{Example_BoW}.

Another common approach to text vectorization in text analysis is known as “Term Frequency – Inverse Document Frequency” (TF-IDF) method. 
The Inverse Document Frequency (\citeauthor{Jones72astatistical}, \citeyear{Jones72astatistical}) is a scoring of how rare is a word across documents:
\begin{equation*}
    \IDF(w_j) = \log \frac{n}{n_j},
\end{equation*}
where $n$ is the number of documents in the corpus, and $n_j$ is the number of documents where the word $w_j$ appears. 
In TF-IDF (\citeauthor{10.1145/280765.280786}, \citeyear{10.1145/280765.280786}), 
the value of the word $w_j$ in the document $d_i$ is given by 
\begin{equation*}
    \TFIDF(d_i,w_j) = \TF(d_i,w_j) \times \IDF(w_j), 
\end{equation*}
where $i=1, \dots, n$ and $j=1, \dots, m$, and $m$ is the dictionary size.

The similarity between documents can be calculated using the “cosine similarity".
Considering the documents $d_s$ and $d_t$, $s, t \in \{1, \dots, n\}$, represented by 
\begin{equation*}
    \hskip -0.5cm \boldsymbol{x}_s = \left( \TFIDF(w_1,d_s), \dots, \TFIDF(w_m,d_s) \right) \text{,}
\end{equation*}
\begin{equation*}
    \boldsymbol{x}_t = \left( \TFIDF(w_1,d_t) , \dots, \TFIDF(w_m,d_t) \right) \text{,}
\end{equation*}
their cosine similarity (\citeauthor{Singhal_2001}, \citeyear{Singhal_2001}) is given by the cosine of the angle between the two vectors representing the two descriptions:
\begin{equation*}
    \CS(d_s, d_t) = \frac{\boldsymbol{x}_s \cdot \boldsymbol{x}_t}{\lVert \boldsymbol{x}_s \rVert \lVert \boldsymbol{x}_t \rVert}  \in [0,1] \text{.}
\end{equation*}

\subsection{Semantic adjustment}
\label{SemanticStemming}

\begin{figure}
    \centering
    \includegraphics[width=14cm]{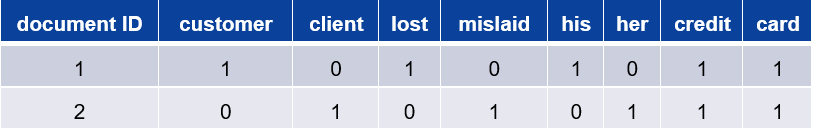}
    \caption{TF document matrix.}
    \label{TF-IDF}
\end{figure}

The TF and TF-IDF approaches alone are not able to capture semantic information, such as the semantic similarity between synonyms. 
In fact, even if two documents are almost identical in terms of meaning, the similarity between them on the basis of TF or TF-IDF could be low due to scarce word matching.
In the following example, we compare two descriptions:
\begin{enumerate}
    \item “The customer lost his credit card"
    \item “The client mislaid her credit card"
\end{enumerate}
The TF matrix 
is reported in Figure \ref{TF-IDF} (stop-word “the" has been removed).
Cosine similarity between the descriptions can be calculated as
\begin{equation*}
    \CS(d_1, d_2) = \frac{\boldsymbol{x}_1 \cdot \boldsymbol{x}_2}{\lVert \boldsymbol{x}_1 \rVert \lVert \boldsymbol{x}_2 \rVert} 
    = \frac{(1,0,1,0,1,0,1,1) \cdot (0,1,0,1,0,1,1,1)}{\sqrt{5}\sqrt{5}} = \frac{2}{5}
    =0.4 \text{.} 
\end{equation*}
Even if the documents are almost identical, $\CS$ is low due to the poor word overlap. 
The columns “customer”, “lost” and “his” should be correlated respectively with the value of columns “client”, “mislaid” and “her”, since they represent the same concepts.
To consider semantic similarity, an adjustment can be applied to the document-by-term matrix using word embedding techniques, such as \texttt{word2vec} (\citeauthor{DBLP:journals/corr/abs-1301-3781}, \citeyear{DBLP:journals/corr/abs-1301-3781}).

\texttt{Word2vec} is built on a neural network-based algorithm to represent words in a vector space, so that different words that share a common concept are “close" as measured by cosine similarity.
Therefore, the cosine similarity between words represents a measure of semantic similarity between them.
    
\begin{figure}
    \centering
    \includegraphics[width=14cm]{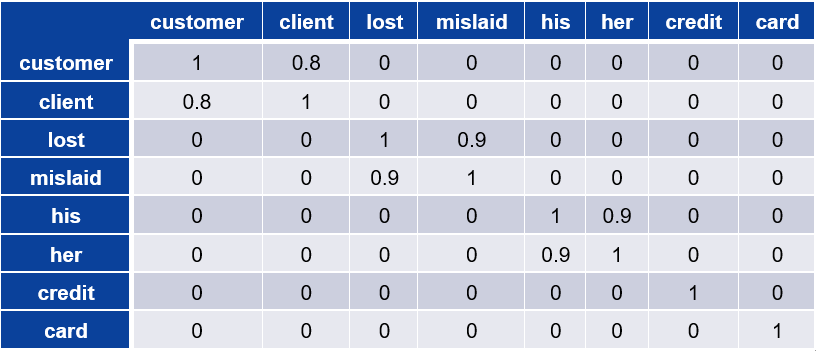}
    \caption{Word similarity matrix.}
    \label{WordSimilarity}
\end{figure}

For example, assume that the word-similarity matrix, shown in Figure \ref{WordSimilarity}, is obtained by applying \texttt{word2vec}. 
The word similarity matrix allows to update the value of each “zero” of document-by-term matrix with the value of the most similar word included in the same row of the document-by-term matrix and scaled by the respective word similarity score (\citeauthor{63e7930434f0418d9183c58093048a6d}, \citeyear{63e7930434f0418d9183c58093048a6d}). 

\begin{figure}
    \centering
    \includegraphics[width=14cm]{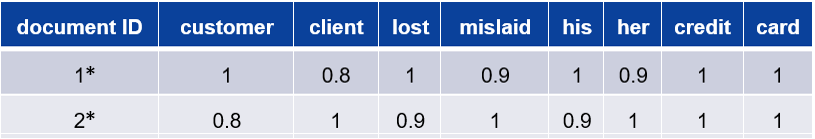}
    \caption{Semantic-aware document-by-term matrix.}
    \label{SemanticTF-IDF}
\end{figure}

The resulting semantic-aware document-by-term matrix is reported in Figure \ref{SemanticTF-IDF}.
The cosine similarity between the two documents can now be recalculated on the basis of the semantic-aware document-by-term matrix:
\begin{equation*}
    \hskip -0.5cm 
    \CS(d_{1^*}, d_{2^*}) =
    \frac{(1,0.8,1,0.9,1,0.9,1,1) \cdot (0.8,1,0.9,1,0.9,1,1,1)}{\sqrt{7.26}\sqrt{7.26}} 
    = 0.992 \text{.} 
\end{equation*}
The similarity score between the two documents, considering the semantic adjustment, increases from 0.4 to around 0.99, reflecting the actual similarity between them.

\subsection{Dimensionality reduction}

\begin{figure}
        \centering
        \includegraphics[width=12cm]{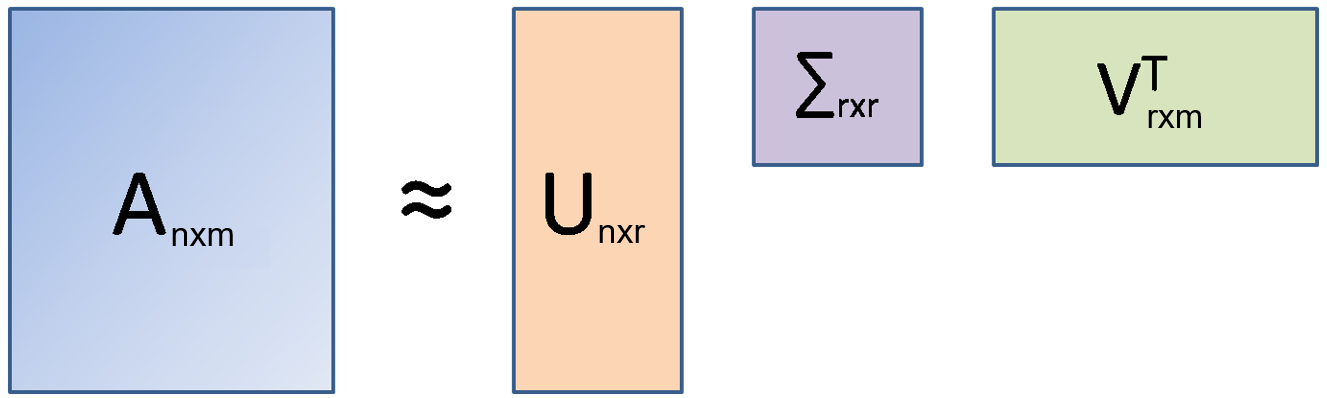}
        \caption{Singular value decomposition (SVD) representation.}
        \label{SVD}
    \end{figure}

After introducing a semantic measure of similarity to extract information from texts, the next step of the proposed workflow is to identify clusters of similar descriptions. 
Convenient approaches make use of 
dimensionality reduction methods, used to map document vectors from the word space 
to a space whose reduced dimensionality is user-defined. 
The Latent Semantic Analysis (LSA) (\citeauthor{Dumais_1996}, \citeyear{Dumais_1996}) is based on the Singular Value Decomposition (SVD) in which the document-by-term matrix $A$ (see Figure \ref{SVD}) is reduced to a set of orthogonal factors from which the original matrix can be approximated. 
Multidimensional projection techniques such as Least Square Projection (LSP) (\citeauthor{4378370}, \citeyear{4378370}) can also be adopted to preserve neighborhood relations.
Since the similarity between documents can still be measured in the reduced space represented by the matrix $U$ (see Figure \ref{SVD}), text objects can be ranked by their similarity.
For example, by identifying a point in the space (representing an event description), the text objects in its neighborhood can be identified. 

\subsection{Cluster selection}
\label{Cluster}

\begin{figure}
    \centering
    \includegraphics[width=10cm]{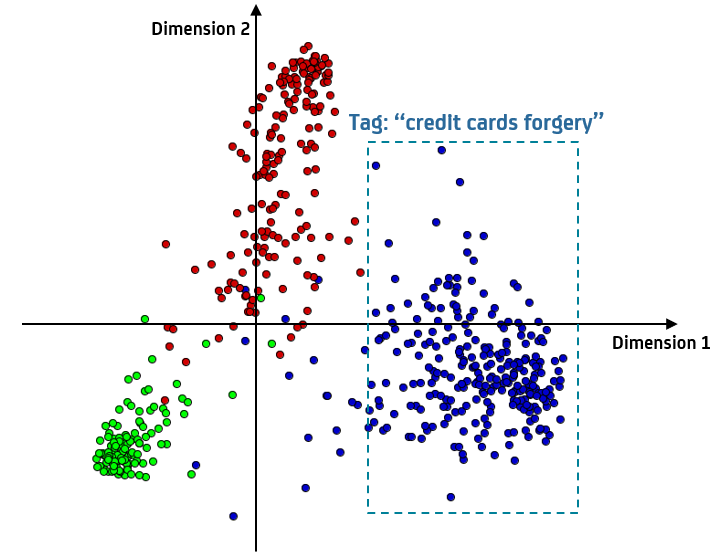}
    \caption{A graphical representation showing document similarities in the 2D space: each point represents a document, and each color represents a document cluster.}
    \label{Similarity2DSpace}
\end{figure}

 Dimensionality reduction is used to build a 2D representation of the data (\citeauthor{info9040100}, \citeyear{info9040100}), 
where each point is an event description and similar ones are represented as a cluster of points. 
Using the 2D representation, the analysts can explore a large volume of documents, identifying clusters of similar documents as groups of points close to each other,
to understand their content and assign tags, as “credit card forgery”, as reported in the example of Figure \ref{Similarity2DSpace}.
 For example, we suppose that most of the blue points can be tagged by the analysts as “credit card forgery”.


\subsection{Cluster validation}
\label{ClusterValidation}

Once the analysts have tagged the events belonging to the identified clusters, it is possible to apply statistical clustering and classification techniques to validate them.
These techniques can be also applied to support analysts' activity when there is a huge amount of data to be tagged.
There are several methods that can be adopted for this task.
We have identified the following approaches among the most used and popular ones:
\begin{itemize}
    \item $k$-means clustering (\citeauthor{Macqueen67somemethods}, \citeyear{Macqueen67somemethods}), where the data are partitioned into $k$ groups, such that the sum of the squared Euclidean distances between the points and centers of the assigned clusters is minimized.
    The algorithm, starting from initial $k$ centers, iterates until convergence by recalculating the centers of clusters and reassigning the points to the clusters on the basis of distances.
    Since the initial $k$ centers are randomly selected, it is a good practice to rerun the algorithm with several initializations in order to select the best clustering among the selected ones.
    $k$-means is implemented in the function \texttt{kmeans} of the R programming language.
    The quality of the obtained clustering can be assessed using the silhouette plots (\citeauthor{ROUSSEEUW198753}, \citeyear{ROUSSEEUW198753}).
    For each data point, a silhouette value is calculated (and plotted), measuring how similar it is to its own cluster (cohesion) compared to the other clusters (separation). 
    This value belongs to the range $[-1, +1]$, where a high value indicates that the point is well matched to its cluster and poorly matched to other clusters. 
    If most points have high values, then the obtained clustering is appropriate. 
    The average value, called silhouette index, is usually adopted as a synthetic index of clustering quality.
    \item Spherical $k$-means clustering, which is based on cosine distance (\textit{i.e.}, 1 minus the cosine similarity) instead of the Euclidean distance.
    Note that this method is equivalent to scaling data to unit length, and then using standard $k$-means.
    This method is suggested to mitigate the effect of different document lengths (\citeauthor{Dhillon_2001}, \citeyear{Dhillon_2001}), and it is implemented in the R package \texttt{skmeans} (\citeauthor{JSSv050i10}, \citeyear{JSSv050i10}).
    \item Clustering via Gaussian finite mixture models implemented in the R package \texttt{mclust} (\citeauthor{Scrucca_2016}, \citeyear{Scrucca_2016}).
    \item Trimmed $k$-means clustering implemented in the R package \texttt{tclust} (\citeauthor{Fritz_2012}, \citeyear{Fritz_2012}).
    In particular, the trimmed $k$-means is obtained by the function \texttt{tclust}, setting 1 as restriction factor on eigenvalues (\textit{i.e.}, the ratio between the maximal and minimal eigenvalues).
    \item Mixtures of Unigrams described by \citeauthor{NigamK_2000} (\citeyear{NigamK_2000}) and implemented by the function \texttt{mou\_EM} in the R package \texttt{DeepMOU} (\citeauthor{Viroli_2021}, \citeyear{Viroli_2021}).  
    Parameter estimation is performed by means of the Expectation-Maximization (EM) algorithm.   
    \item Deep Mixtures of Unigrams described by \citeauthor{Viroli_2020} (\citeyear{Viroli_2020}) and implemented by the function \texttt{deep\_mou\_gibbs} in the R package \texttt{DeepMOU}.
    Parameter estimation is performed by means of Gibbs sampling. 
    \item Dirichlet-Multinomial Mixtures model described by \citeauthor{Anderlucci_2020} (\citeyear{Anderlucci_2020}) and implemented by the function \texttt{dir\_mult\_GD} in the R package \texttt{DeepMOU}.
    Parameter estimation is performed by means of a Gradient Descend algorithm.
    \item Latent Dirichlet Allocation (LDA) is a generative statistical model that explains a set of observations through unobserved groups. 
    It is an example of a topic model, where observations (\textit{e.g.}, words) are collected into documents. 
    It assumes that the words in a document are drawn from $k$ topics, and each topic is characterized by a probability distribution over the available words. 
    Each document is supposed to contain a small number of topics.
    An application of LDA in the context of text mining is described by \citeauthor{Blei2003} (\citeyear{Blei2003}).
    LDA is implemented by the function \texttt{FitLdaModel} in the R package \texttt{textmineR} (\citeauthor{Jones_2021}, \citeyear{Jones_2021}).
    Parameter estimation is performed by means of Gibbs sampling. 
\end{itemize}
The consistency between the cluster selection, performed as described in Section \ref{Cluster}, and the results of the aforementioned approaches can be assessed through the accuracy measure, which is calculated as follows for a classification method:
\begin{equation*}
    \text{Accuracy} = \frac{\text{Number of correctly classified data}}{\text{Total number of data}} \text{.} 
\end{equation*}
Several other measures can be used to assess performances of classification methods (\textit{e.g.}, precision, recall, F1 score), but accuracy appears to be the most intuitive and sufficiently general to be applied for the aforementioned approaches.

\section{Application to operational risk data}
\label{Application}

The objective of this application is to analyse the descriptions of the CoRep Operational Risk data set for UniCredit banking group using all the approaches described in the previous sections.
The CoRep is the Common Reporting, which is the set of all data that the financial institutions have to periodically report to their Supervisory Authorities (\textit{e.g.}, European Central Bank). 
Among the CoRep reports, there is the C17.02 template, which reports information (including the description) on Operational Risk events leading to gross loss amounts higher than or equal to \euro $\,$ 100,000 (gross means without considering any recovery). 
The analyzed data set is composed of the OpRisk data which are relevant for the C17.02 template, considering that this template has been in place since 2018. 
Each record of this data set represents an OpRisk event, 
while the main fields report the following data:
\begin{itemize}
    \item Event ID: the ID of the Operational Risk event 
    \item Date of Accounting: the first accounting date of the event 
    \item Event Type: the Basel Event Type level 1 classification of the event 
    \item Gross Loss: the total gross loss amount in \euro $\,$ for the event,
    \textit{i.e.}, the sum of economic impacts related to the event: 
    losses, provisions and releases of provisions
    \item Description: the description of the event, which is a text field reporting an English anonymized description, having a maximum of 250 characters 
\end{itemize}
This application concerns the part of CoRep Operational Risk data set related to the OpRisk events having event type “Clients, Products \& Business Practices" and first accounting date between 2018 and 2021.
This selection leads to a data set composed of 644 events with relevant descriptions.

The analysis is performed using the R packages \texttt{quanteda} (\citeauthor{Benoit_2018}, \citeyear{Benoit_2018}), \texttt{word2vec} (\citeauthor{Wijffels_2021}, \citeyear{Wijffels_2021}), and the ones mentioned in Section \ref{ClusterValidation}.

First of all, the descriptions are cleaned as described in Section \ref{Cleaning}.
There is no need for language splitting and data anonymization, since such descriptions are all entered in English language and without any personal information. 
The stop-word list, already specified in Section \ref{Workflow}, has been obtained through the R package \texttt{stopword} (\citeauthor{Benoit_2021},  \citeyear{Benoit_2021}).

Since the analyzed descriptions are short texts (having max 250 characters), we apply the TF weighting schema without any IDF scaling, as motivated by \citeauthor{anderlucci2019unknown} (\citeyear{anderlucci2019unknown}) for their application on similarly structured data.
We point out that the main purpose of the IDF scaling is to reduce the weight of terms that are used in many documents under the hypothesis that if a word is used in many descriptions, then it is not informative, and then it is not useful to discriminate different clusters of data.
However, most of the non-informative terms have been already excluded by removing the stop-words from the text corpus.
Therefore, applying the IDF scaling to short texts, we risk reducing the weights of some informative words which characterize the clusters.
This aspect is verified during the next steps of the analysis.

We obtain a document-by-term matrix having 644 rows (\textit{i.e.}, the number of descriptions) and 1037 columns (\textit{i.e.}, the length of the dictionary consisting of all the unique words included in the cleaned descriptions).

\begin{figure}
        \centering
        \includegraphics[width=14cm]{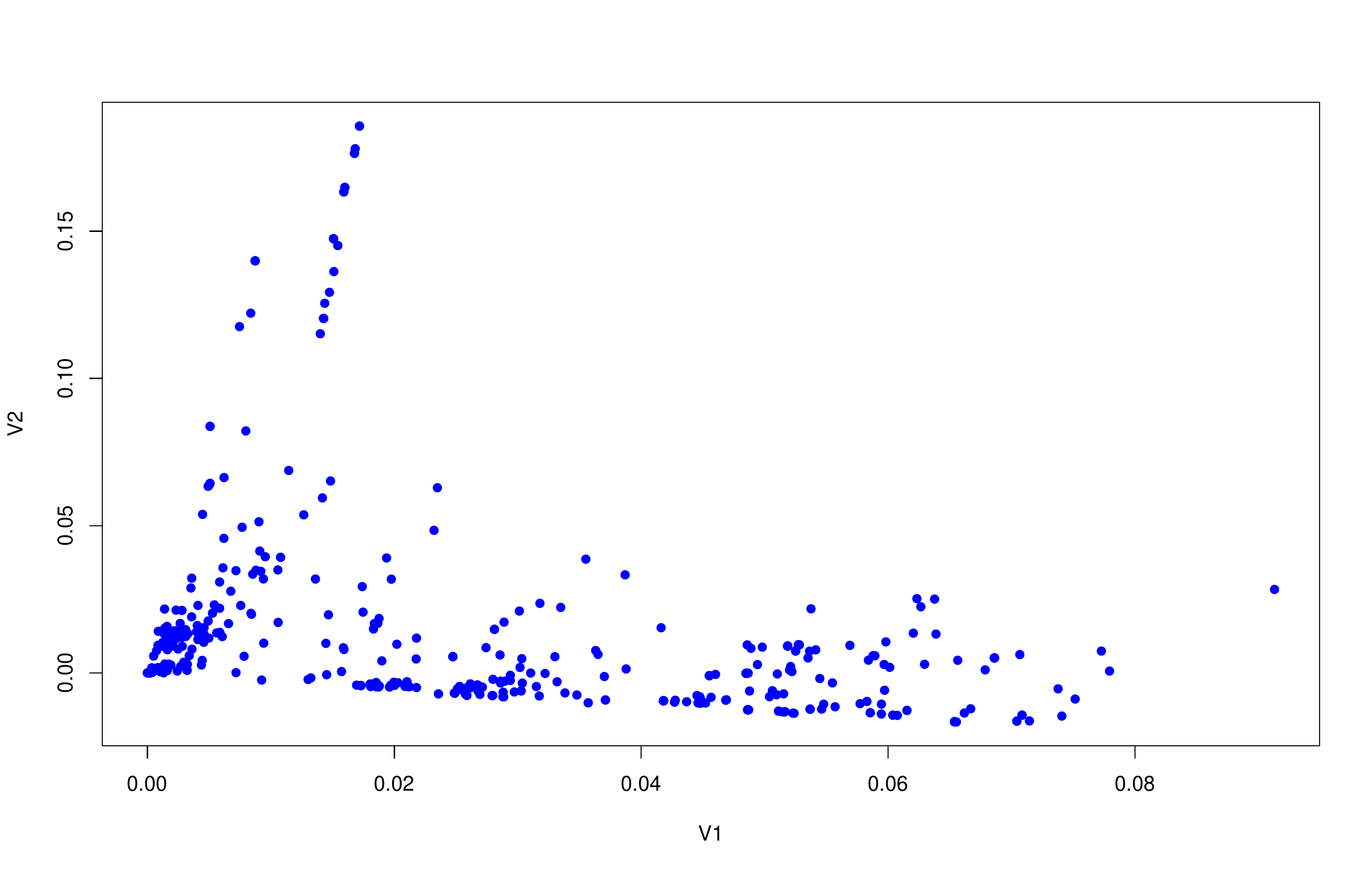}
        \caption{2D representation of the document-by-term matrix}
        \label{CoRep-ET4-TF}
    \end{figure}

We apply the LSA to the document-by-term matrix to obtain a 2D representation, reported in Figure \ref{CoRep-ET4-TF}, where the axes $V1$ and $V2$ represent the first two LSA dimensions.

\begin{figure}
    \centering
    \includegraphics[width=14cm]{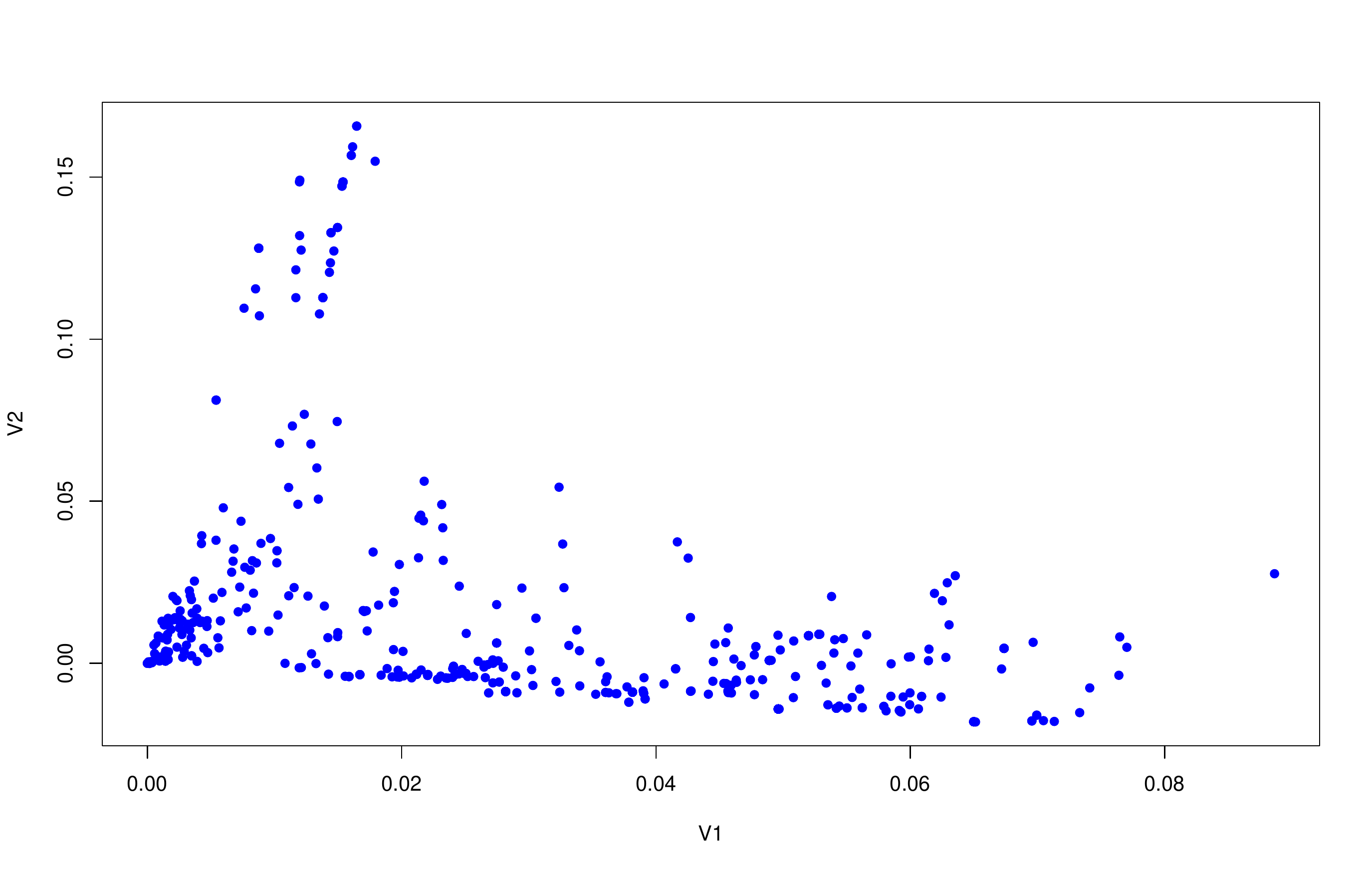}
    \caption{2D representation of the semantic-aware document-by-term matrix.}
    \label{CoRep-ET4-SemanticTF}
\end{figure}

The next step is to generate the semantic-aware document-by-term matrix using the approach described in Section \ref{SemanticStemming}.
We use a  pre-trained word embedding obtained by the \texttt{word2vec} approach (available at \citetitle{NLPL} (\citeyear{NLPL}), selecting ID=40, \textit{i.e.}, "English CoNLL17 corpus").
This allows us to obtain the word similarity matrix by calculating the cosine similarity between all the pairs of words included in the dictionary of the data set.
The word similarity matrix is then used to adjust the document-by-term matrix, as described in Section \ref{SemanticStemming}.
Similarly to \citeauthor{63e7930434f0418d9183c58093048a6d} (\citeyear{63e7930434f0418d9183c58093048a6d}), we use a similarity matrix that only contains similarity values higher than 0.8 in order to avoid including noise (\textit{i.e.}, medium-low similarity due more to randomness than similar meaning) into the semantic adjustment.
Some rationales for the selection of threshold 0.8 are reported in the next steps of the analysis.
Applying the LSA, the obtained semantic-aware document-by-term matrix is represented in 2D in Figure \ref{CoRep-ET4-SemanticTF}.

\begin{figure}
        \centering
        \includegraphics[width=14cm]{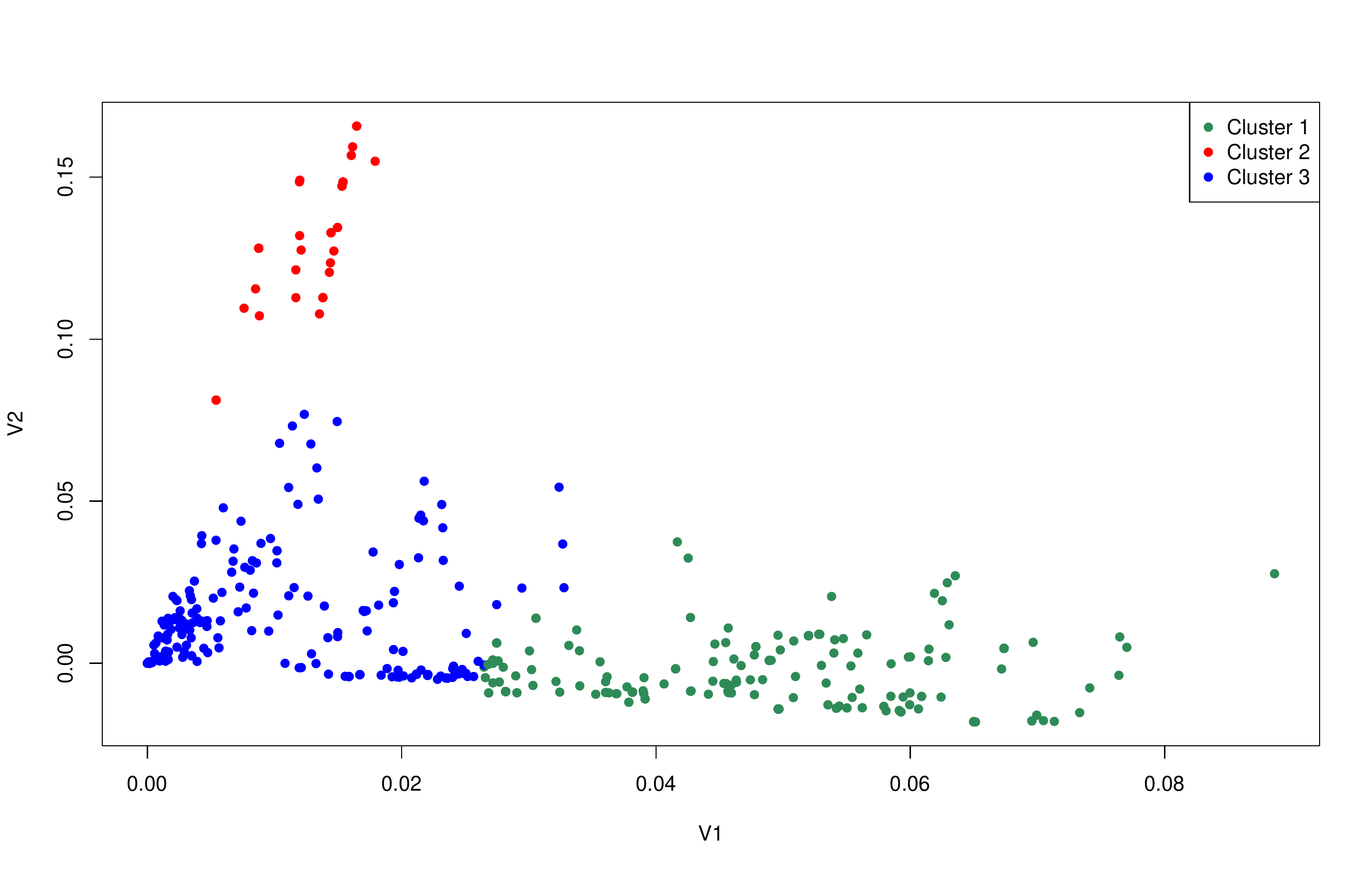}
        \caption{2D representation of the semantic-aware document-by-term matrix with the identified clusters, with similarity threshold 0.8: 
        cluster 1 represents “disputes related to irregularities in the interest rates calculation”, cluster 2 identifies “disputes related to mortgages in foreign currency”, and other events are in cluster 3.}
        \label{CoRep-ET4-SemanticTF-Clusters}
    \end{figure}

The plot supports the activity of the analysts, who decided to tag, after having examined the closer points, two clusters of events (clusters 1 and 2 in Figure \ref{CoRep-ET4-SemanticTF-Clusters}), and a cluster of residual events (cluster 3 in Figure \ref{CoRep-ET4-SemanticTF-Clusters}).
There are some common-meaning words, identified by the analysts, that are contained in all descriptions within clusters 1 and 2.
Based on these common-meaning words, it emerges that the two identified clusters, representing two different root-causes for OpRisk, and the residual cluster, can be described as follows:
\begin{enumerate}
    \item Disputes related to irregularities in the interest rates calculation (composed of 384 events)
    \item Disputes related to mortgages in foreign currency (composed of 48 events)
    \item Other events  (composed of 212 events)
\end{enumerate}

As it can be seen by comparing Figures \ref{CoRep-ET4-TF} and \ref{CoRep-ET4-SemanticTF}-\ref{CoRep-ET4-SemanticTF-Clusters}, the semantic-aware document-by-term matrix allows to include into the clusters also descriptions expressing similar concepts, even when they do not include the same significant words identifying the clusters.  

\begin{figure}
        \centering
        \includegraphics[width=14cm]{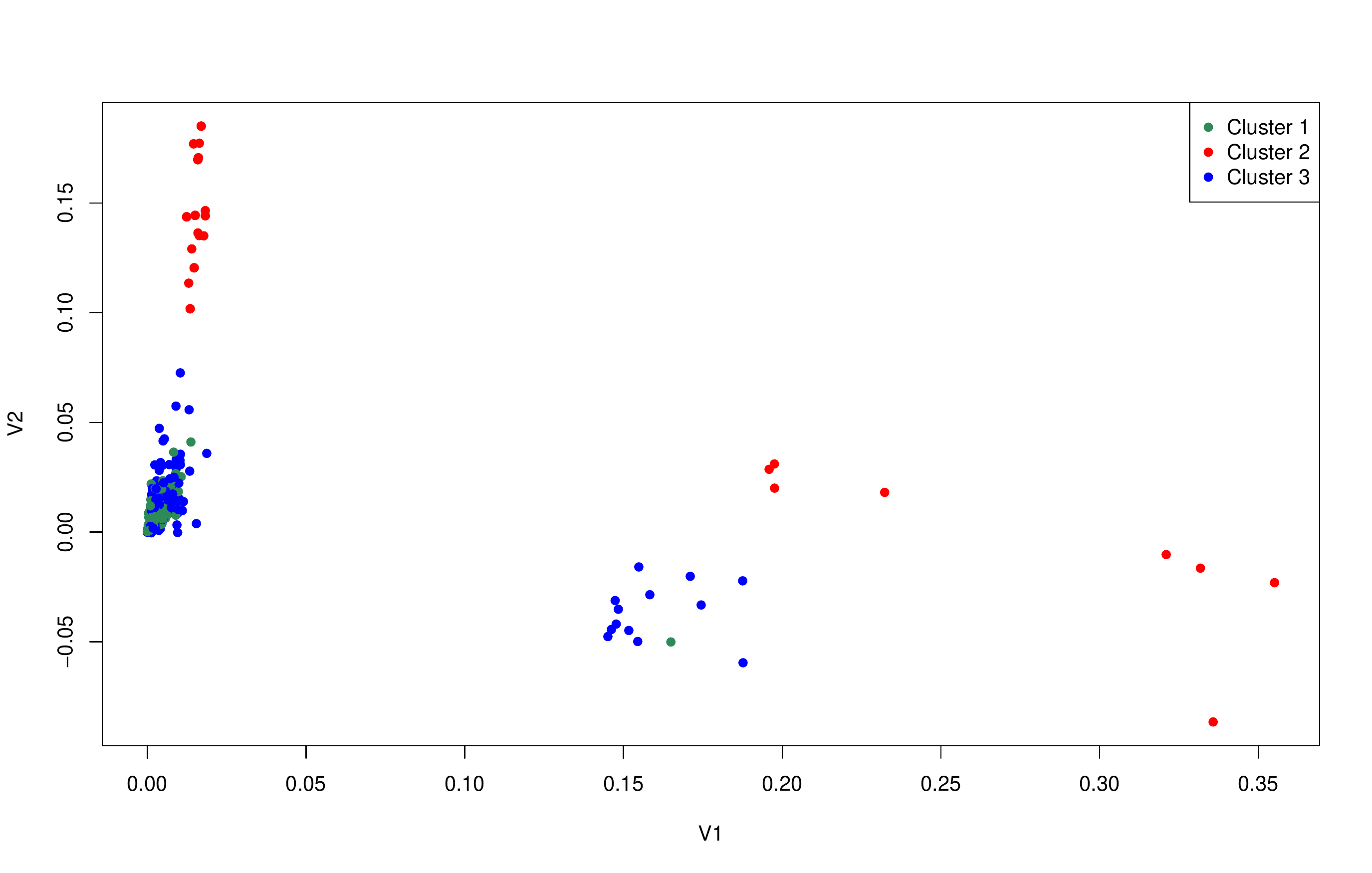}
        \caption{2D representation of the semantic-aware TF-IDF matrix with the identified clusters:
        cluster 1 represents “disputes related to irregularities in the interest rates calculation", cluster 2 identifies “disputes related to mortgages in foreign currency", and other events are in cluster 3.}
        \label{CoRep-ET4-SemanticTFIDF-Clusters}
    \end{figure}

In order to further motivate the exclusion of the IDF scaling, we report in Figure \ref{CoRep-ET4-SemanticTFIDF-Clusters} the chart related to the 2D representation of the semantic-aware TF-IDF matrix (\textit{i.e.}, the TF-IDF matrix including the semantic adjustment based on similarity threshold 0.8) with clusters previously identified by the analysts. 
From Figure \ref{CoRep-ET4-SemanticTFIDF-Clusters}, it appears a significant overlap between clusters 1 and 3.
In fact, since events related to cluster 1 are identified by a few words that are basically included in all its descriptions, the IDF scaling significantly reduces the weights of such terms, moving most of the related points very close to the chart origin (\textit{i.e.}, very close to the point $(V1=0, V2=0)$ in the chart).
This representation would make, for the analysts, the task of distinguishing between cluster 1 (disputes related to irregularities in the interest rates calculation) and the residual cluster quite hard.



\begin{figure}
        \centering
        \includegraphics[width=14cm]{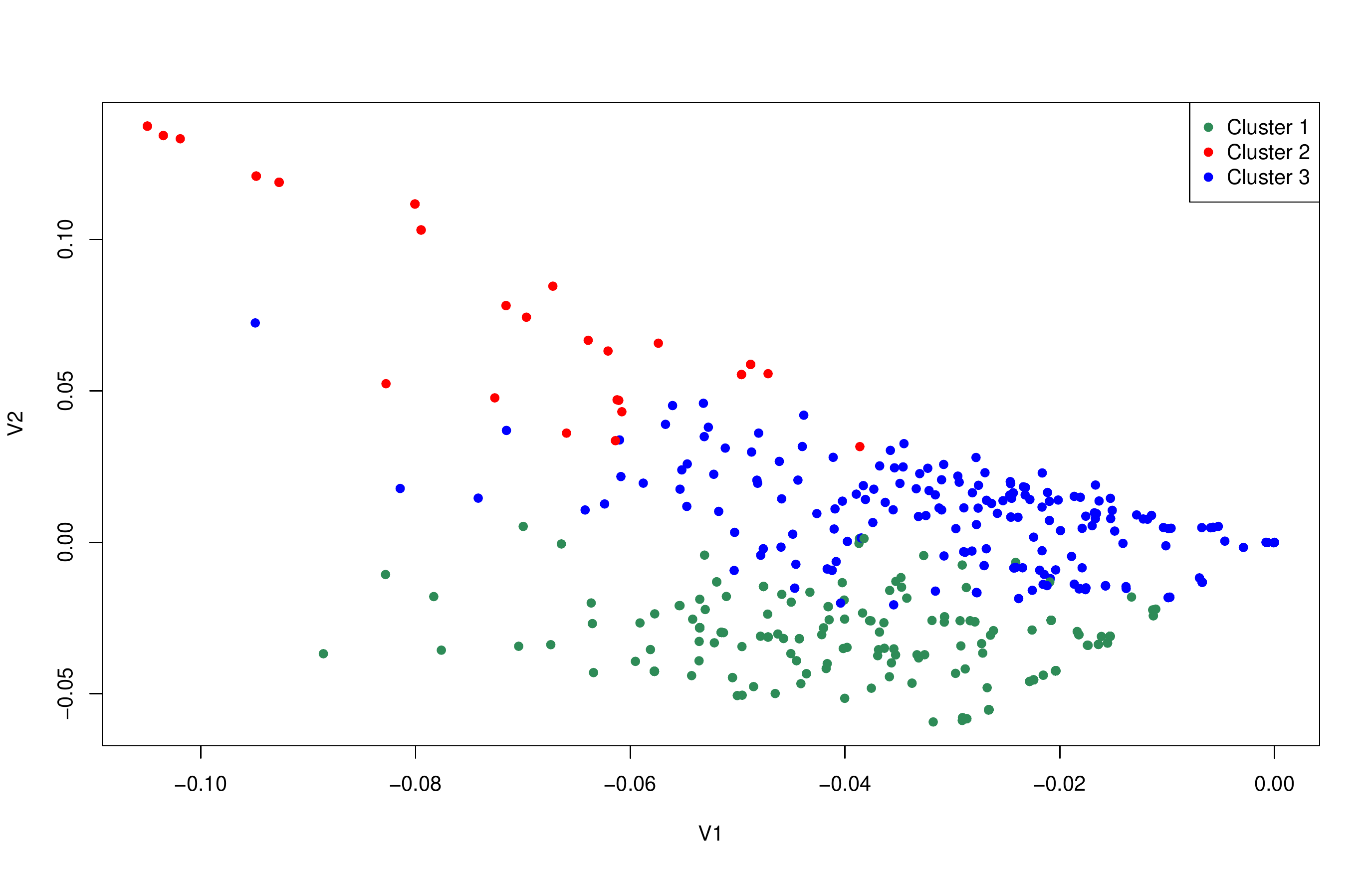}
        \caption{2D representation of the semantic-aware document-by-term matrix with the identified clusters, considering similarity threshold 0.7.}
        \label{CoRep-ET4-SemanticTF-Thr07-Clusters}
    \end{figure}

\begin{figure}
        \centering
        \includegraphics[width=14cm]{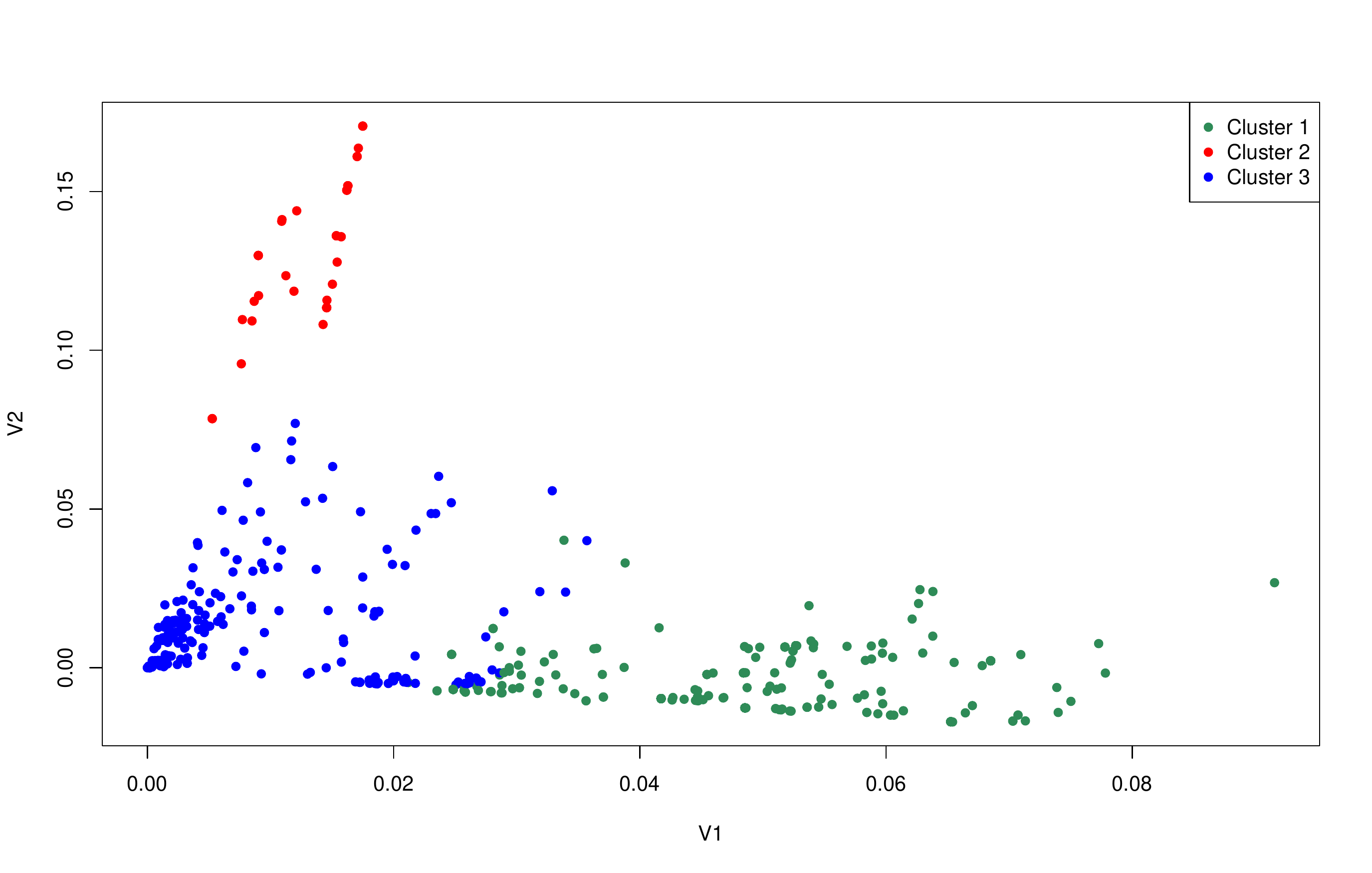}
        \caption{2D representation of the semantic-aware document-by-term matrix with the identified clusters, considering similarity threshold 0.9.}
        \label{CoRep-ET4-SemanticTF-Thr09-Clusters}
    \end{figure}

In order to motivate the selection of similarity threshold 0.8, we also report the charts obtained with similarity thresholds 0.7 (in Figure \ref{CoRep-ET4-SemanticTF-Thr07-Clusters}) and 0.9 (in Figure \ref{CoRep-ET4-SemanticTF-Thr09-Clusters}).
We can see from Figures \ref{CoRep-ET4-SemanticTF-Thr07-Clusters} and \ref{CoRep-ET4-SemanticTF-Thr09-Clusters} that considering threshold 0.7 completely alters the initial configuration, whereas considering 0.9 leaves the configuration very similar to the non-semantic-aware one (Figure \ref{CoRep-ET4-TF}).
Therefore, we can consider the similarity threshold 0.8 (or, at least, the values within a small neighborhood of 0.8) as the best trade-off between including too much noise (\textit{i.e.}, threshold 0.7) and not including any appreciable semantic adjustment (\textit{i.e.}, threshold 0.9). 

\begin{figure}
        \centering
        \includegraphics[width=14cm]{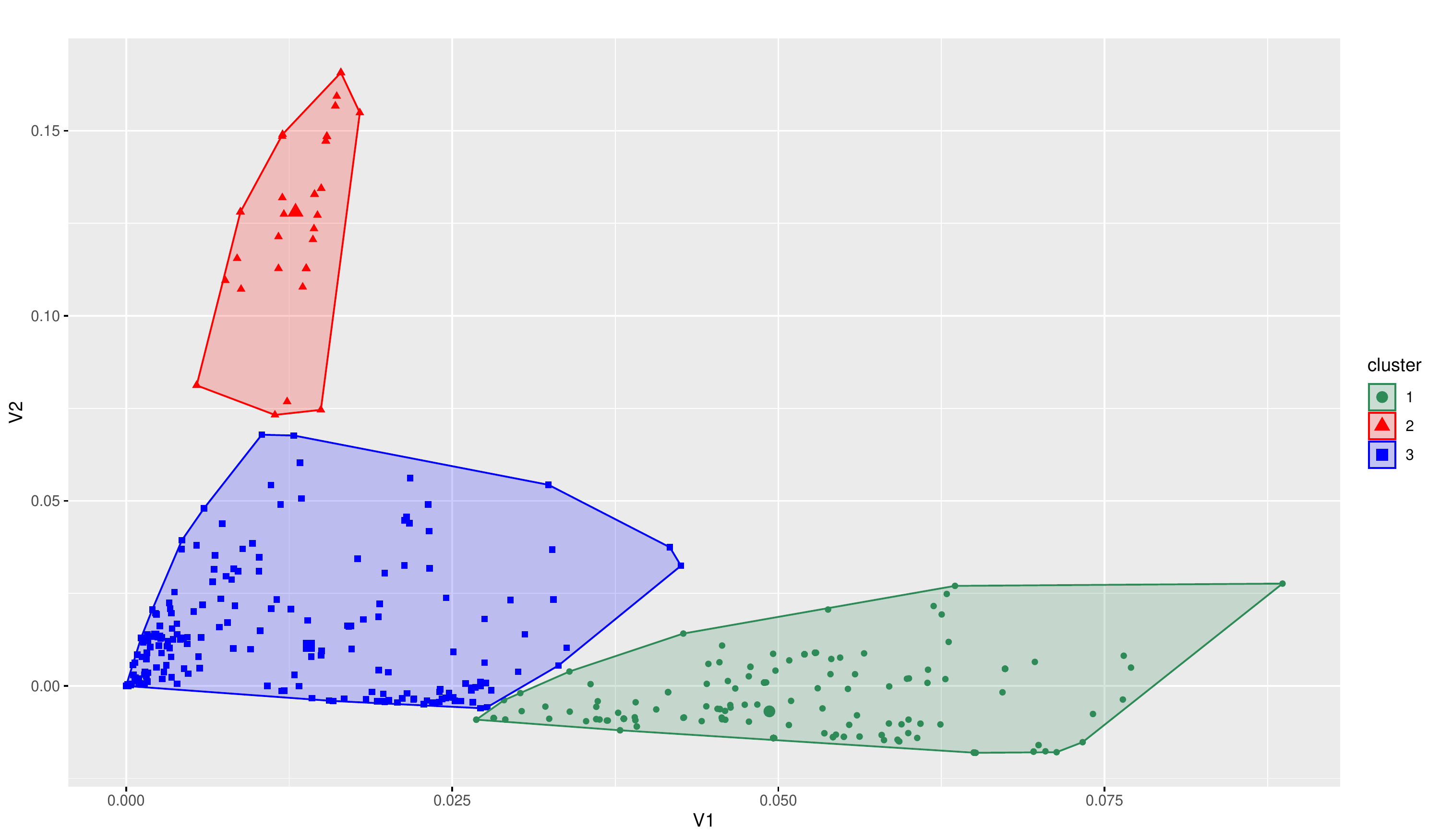}
        \caption{$k$-means clustering with $k=3$ and 1000 starting points.}
        \label{CoRep-ET4-SemanticTF-kmeans}
    \end{figure}

\begin{figure}
        \centering
        \includegraphics[width=14cm]{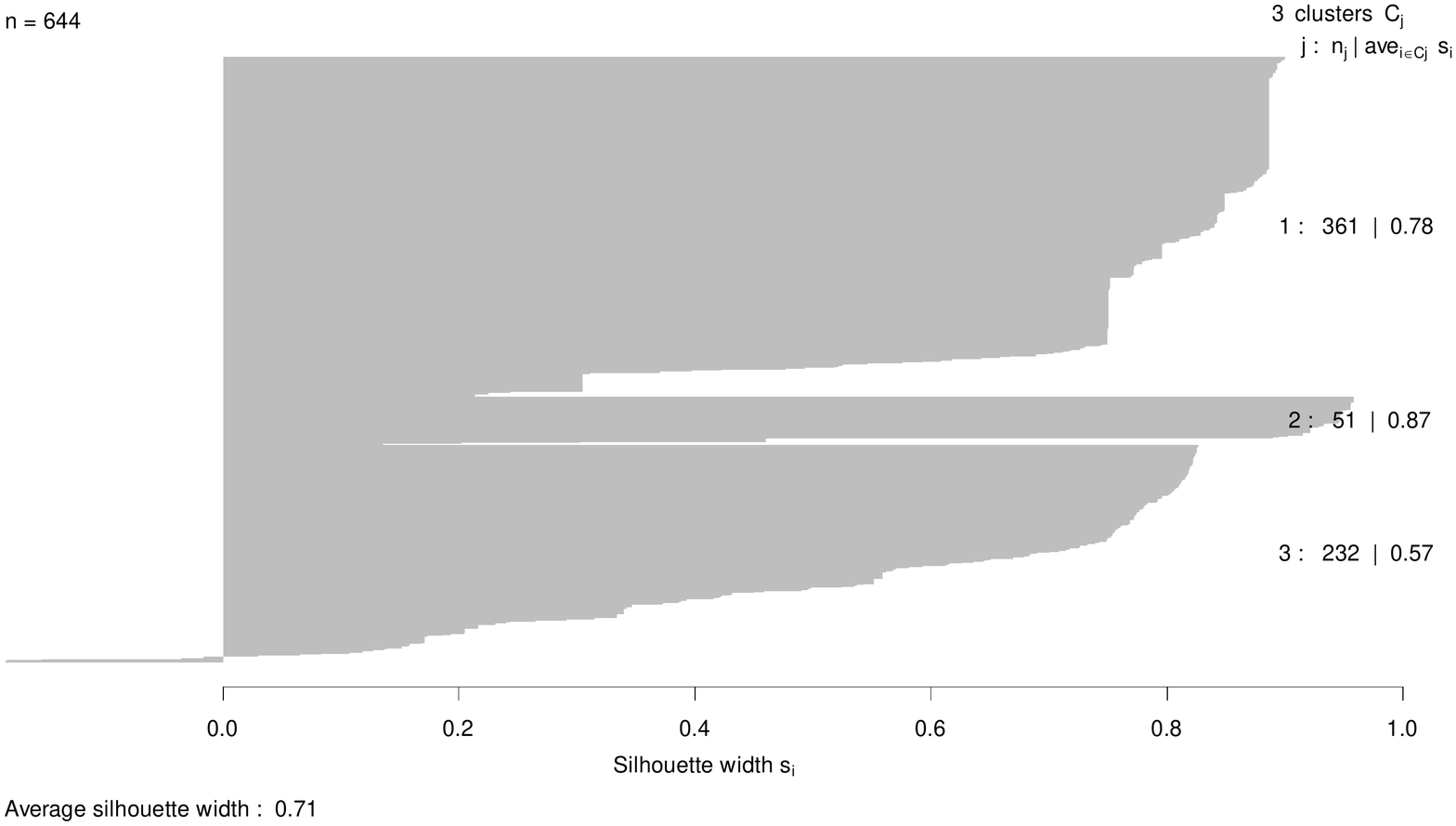}
        \caption{Silhouette plot of the $k$-means clustering.}
        \label{CoRep-ET4-SemanticTF-kmeans-Silhouette}
    \end{figure}

Taking into account the knowledge of the analysts, who identified three clusters (\textit{i.e.}, the two clusters based on common root-cause events and the cluster of residual data), we run a $k$-means clustering 
with $k=3$. 
Moreover, 1000 random starting points are used to avoid being sensitive to a specific starting point selection.
Results are reported in Figure \ref{CoRep-ET4-SemanticTF-kmeans}.
The good quality of the $k$-means clustering is confirmed by the silhouette plot, where the average silhouette index is 0.71 (Figure \ref{CoRep-ET4-SemanticTF-kmeans-Silhouette}).
The $k$-means clustering assigns 361 and 51 events to clusters 1 and 2, respectively. 
Considering 26 misclassified events out of 644 with respect to the analysts selection, we obtain an accuracy 
of around 96\% for the $k$-means clustering.

Other methods described in Section \ref{ClusterValidation} have also been applied on the data.
The accuracies of all considered approaches are reported in Table \ref{TableAccuracy}, including their rankings from the highest to the lowest values.
\begin{table}
\caption{\label{TableAccuracy}Accuracy indexes.}
\centering
\begin{tabular}{c l c} 
 \hline
 \textbf{Rank} & \textbf{Method} & \textbf{Accuracy (\%)} \\ [0.5ex] 
 \hline 
 1 & $k$-means & 95.96 \\ 
 8 & Spherical $k$-means & 61.02  \\
 2 & Gaussian finite mixture models & 95.19  \\
 3 & Trimmed $k$-means & 95.03  \\
 6 & Mixtures of Unigrams & 70.92  \\
 7 & Deep Mixtures of Unigrams & 69.98  \\
 5 & Dirichlet-Multinomial Mixtures & 76.05  \\
 4 & Latent Dirichlet Allocation & 85.40  \\ [1ex] 
 \hline
\end{tabular}
\end{table}
From the reported results, we observe that:
\begin{itemize}
    \item $k$-means clustering, applying 1000 different starting points, 
    shows the highest accuracy (\textit{i.e.}, 96\%).
    
    As an additional motivation for the exclusion of the IDF scaling, the method of $k$-means has also been applied to the semantic-aware TF-IDF matrix, recalculating the first two LSA dimensions. 
    In this case, the accuracy drops to 68\%.
    This significant decrease with respect to the results obtained without the IDF scaling is consistent with Figure \ref{CoRep-ET4-SemanticTFIDF-Clusters} (showing substantial overlapping between clusters 1 and 3).
    
    As a further reason for the selection of the similarity threshold 0.8, we calculated all the accuracy values that we would have obtained by applying $k$-means to the first two LSA dimensions recalculated on the semantic-aware document-by-term matrix based on similarity thresholds between 0.7 and 0.9 (with step 0.05).
    \begin{table}
    \caption{\label{TableAccuracyThreshold}Accuracy indexes for similarity thresholds.}
    \centering
    \begin{tabular}{c c} 
     \hline
     \textbf{Threshold} & \textbf{Accuracy (\%)} \\ [0.5ex] 
     \hline 
     0.70 & 89.29 \\ 
     0.75 & 95.81  \\
     0.80 & 95.96  \\
     0.95 & 90.99  \\
     0.95 & 90.68  \\
     \hline
    \end{tabular}
    \end{table}
    The results reported in Table \ref{TableAccuracyThreshold} confirm that the similarity threshold 0.8 is the best setting also in terms of accuracy.
    \item Spherical $k$-means (\textit{i.e.}, the $k$-means based on the cosine distance) ranked last (\textit{i.e.}, 61\%).
    Even with 10,000 starting points, the accuracy did not substantially improve.
    The poor performance of spherical $k$-means could be due to the similar lengths of analyzed descriptions.
    In this case, the normalization of the vectors representing the descriptions does not seem to be effective in discriminating the correct clusters.
    Intuitively, comparing Figures \ref{CoRep-ET4-SemanticTF-kmeans} and \ref{CoRep-ET4-SemanticTF-skmeans}, it can be noted that the similarity among data is much more due to their Euclidean distance than to the angles between the vectors representing each pair of points.

    \begin{figure}
        \centering
        \includegraphics[width=14cm]{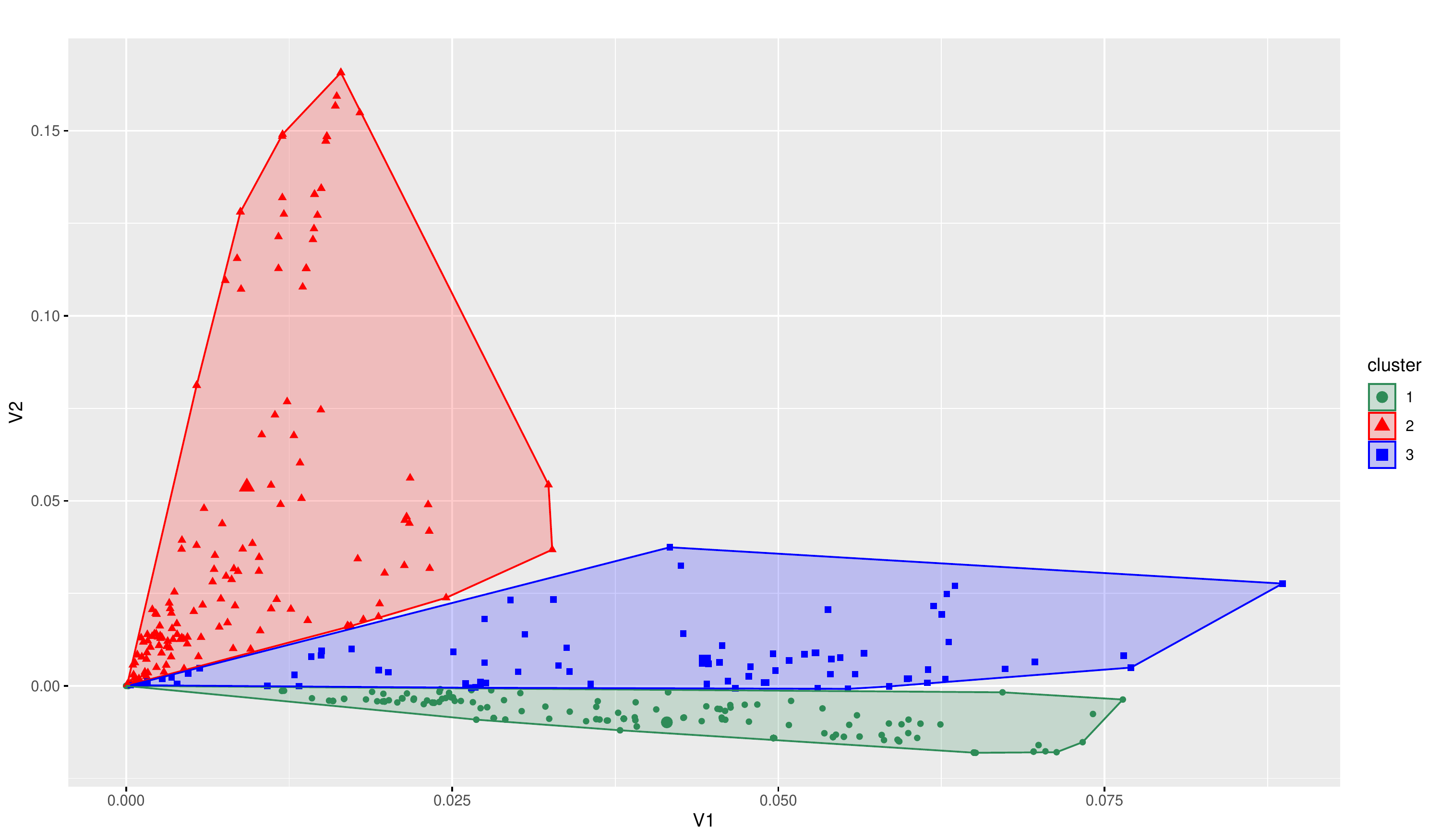}
        \caption{Spherical $k$-means clustering.}
        \label{CoRep-ET4-SemanticTF-skmeans}
    \end{figure}

    \item Gaussian finite mixture models provide a slightly lower accuracy than $k$-means (\textit{i.e.}, 95\%).
    This level of accuracy has been achieved by considering a spherical family with variable volume (\textit{i.e.}, each cluster can include a different number of observations) and equal shape (\textit{i.e.}, each cluster has approximately the same variance so that the distribution is spherical).
    This setting leads to a configuration similar to the one obtained by the $k$-means clustering and, consequently, to a similar accuracy level. 
    The obtained clustering is reported in Figure \ref{CoRep-ET4-SemanticTF-mclust}.

    \begin{figure}
        \centering
        \includegraphics[width=14cm]{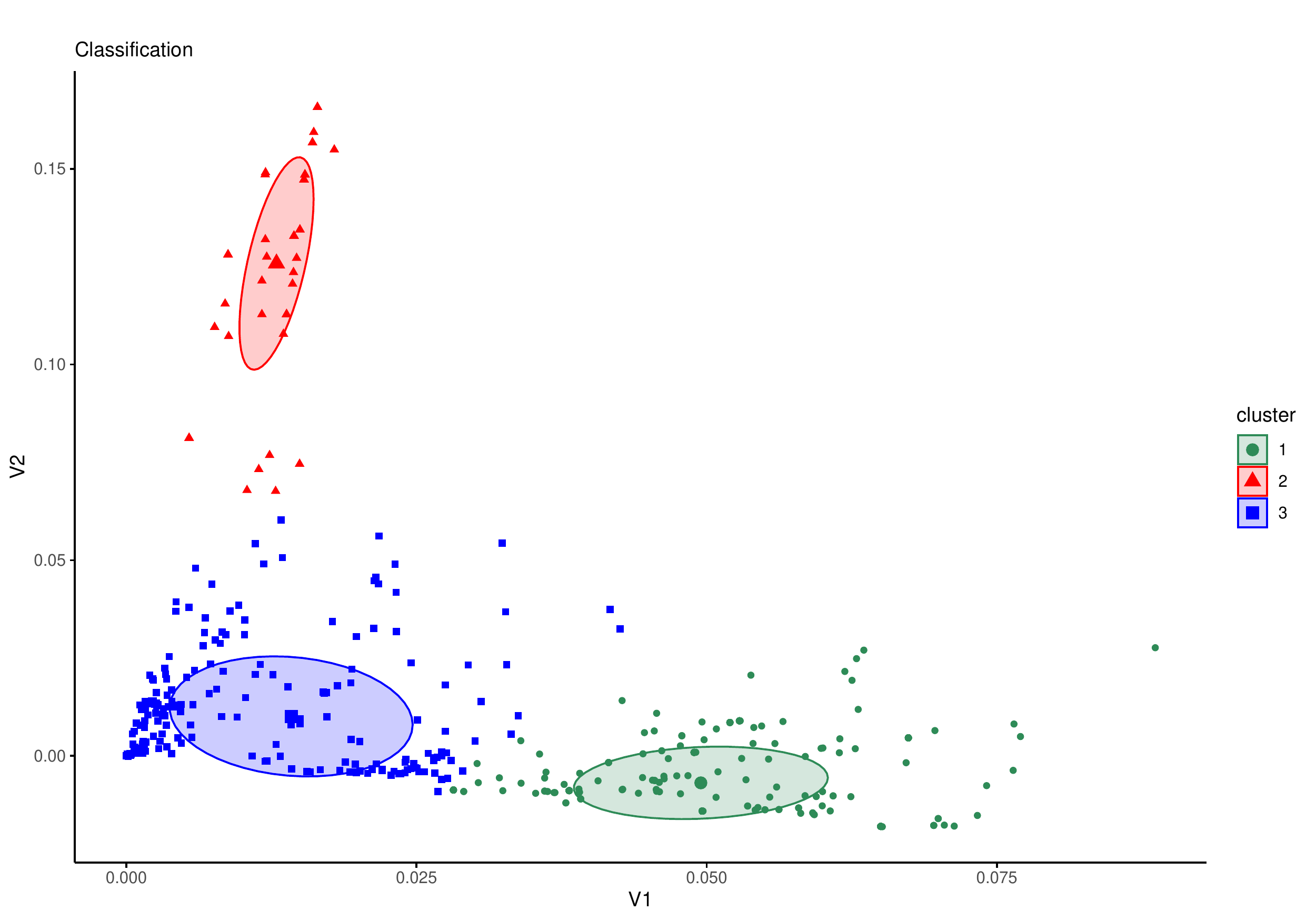}
        \caption{Clustering via Gaussian finite mixture models.}
        \label{CoRep-ET4-SemanticTF-mclust}
    \end{figure}

    \item The method of trimmed $k$-means provides similar accuracy to the Gaussian finite mixture models (\textit{i.e.}, 95\%), again applying 1000 different starting points.
    Different settings have been tested for this method, and the best one (in terms of accuracy) resulted in a proportion of $\alpha=0.02$ trimmed observations (Figure \ref{CoRep-ET4-SemanticTF-tclust}, where the black circles represent the trimmed data, which are not assigned to any cluster).

    \begin{figure}
        \centering
        \includegraphics[width=14cm]{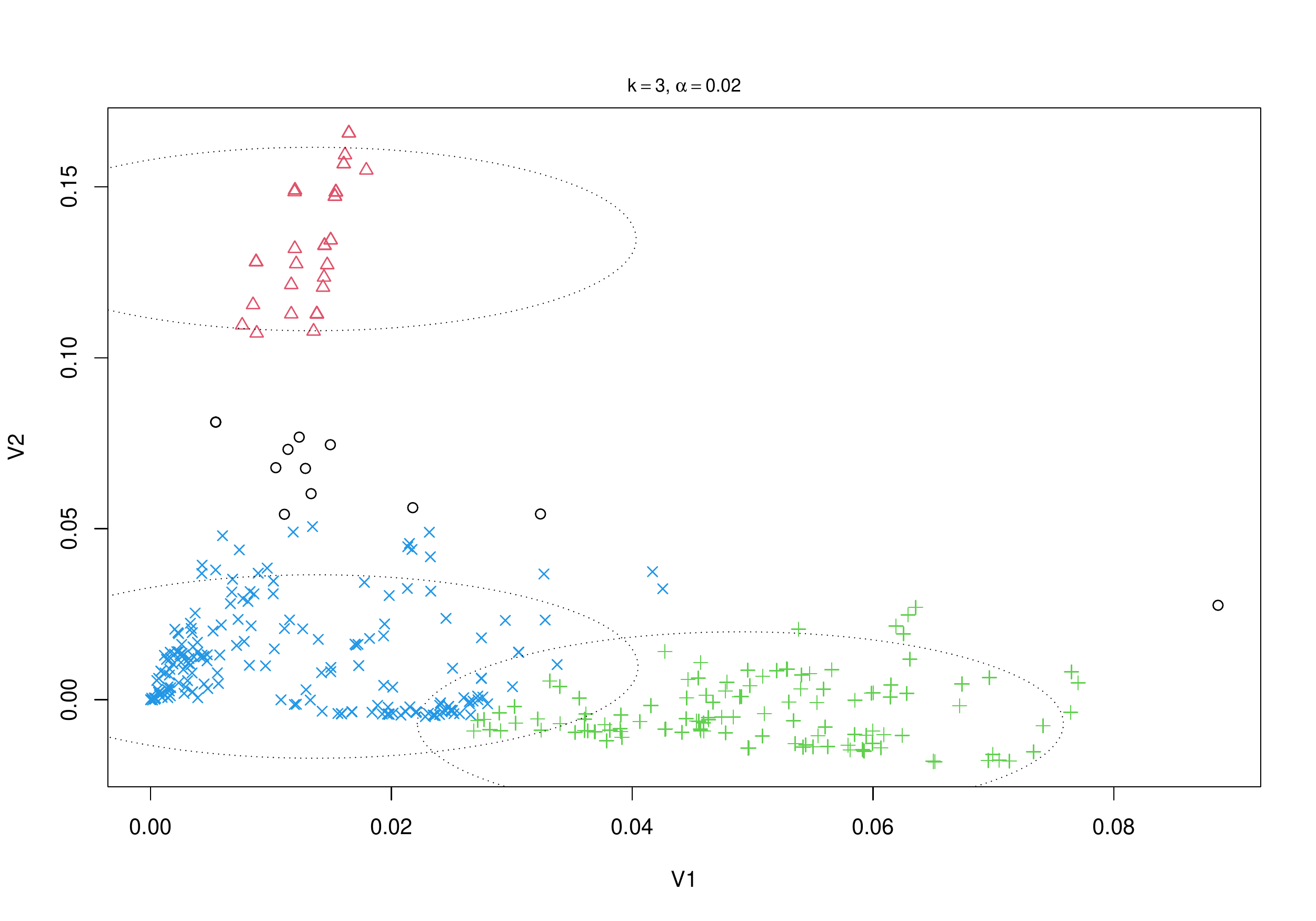}
        \caption{Trimmed $k$-means clustering with $\alpha=0.02$.}
        \label{CoRep-ET4-SemanticTF-tclust}
    \end{figure}

    In order to motivate the choice $\alpha=0.02$, we calculated the accuracy for values of $\alpha$ between 0.01 and 0.1. 
        \begin{table}
    \caption{\label{TableAccuracyAlpha}Accuracy indexes.} 
    \centering
    \begin{tabular}{c c} 
     \hline
     \textbf{$\boldsymbol{\alpha}$} & \textbf{Accuracy (\%)} \\ [0.5ex] 
     \hline 
     0.01 & 91.30 \\ 
     0.02 & 95.03  \\
     0.05 & 92.39  \\
     0.1 & 77.80  \\
     \hline
    \end{tabular}
    \end{table}
    The results, reported in Table \ref{TableAccuracyAlpha}, confirm that the highest accuracy value is obtained for $\alpha=0.02$.
    \item Mixtures of Unigrams, Deep Mixtures of Unigrams, and Dirichlet-Multinomial Mixtures have been directly applied to the document-by-term matrix, adjusted for the semantic similarity, composed of 644 documents and 1037 terms.
    In fact, it is not possible to apply these methodologies to the LSA-based 2D representation, since the LSA can generate negative values, and the three methodologies require as input a matrix composed of positive values.
    Mixtures of Unigrams and Dirichlet-Multinomial Mixtures have been applied using a multi-starting strategy to prevent the local maxima issue, where, for each iteration, the initial assignment to the clusters has been randomly defined. 
    Among all performed iterations, the one having the lowest Bayesian information criterion (BIC) index has been selected, assuring that we have approximately obtained a global maximum value for parameter estimation.
    
    For Mixtures of Unigrams, 1000 iterations (\textit{i.e.}, 10 times the default setting of the function \texttt{mou\_EM}) and a tolerance of $10^{-7}$ (equal to the default setting) have been applied. 
    For the multi-starting strategy, 100 different random starting points have been considered.
    
    For Deep Mixtures of Unigrams, based on Gibbs sampling, 1000 iterations have been used with a burn-in of 500.
    For the top layer, three clusters have been considered, whereas two clusters have been considered for the hidden bottom layer, since this setting provided the highest accuracy in the simulation studies performed by \citeauthor{Viroli_2020} (\citeyear{Viroli_2020}).
    
    For Dirichlet-Multinomial Mixtures, 100 iterations have been set, combined with 100 different random starting points for the multi-starting strategy.
    It is worth mentioning that Dirichlet-Multinomial Mixtures are much more computationally intensive than the Mixtures of Unigrams and the Deep Mixtures of Unigrams, with calculations taking several hours.
    
    These three methods, implemented in the R package \texttt{DeepMOU}, show accuracies between 70\% and 76\%.
    For these approaches, better performances can perhaps be obtained by trying different settings and, in particular, increasing the 
    iterations at the price of higher computational costs.
    \item For the same reason as for the previous methods, the Latent Dirichlet Allocation has also been directly applied to the document-by-term matrix, adjusted for the semantic similarity.
    The applied LDA setting considers three topics (since the analysts identified two clusters, besides the residual data), and 10,000 iterations with a burn-in of 5000.    
    The prior parameters for topics over documents and for words over topics have been set to $\alpha=0.1$ and $\beta=0.05$ (\textit{i.e.}, the default values of the function \texttt{FitLdaModel}).
    To obtain the clustering, each description has been assigned to the topic showing the highest probability.
    The LDA shows an accuracy of around 85\%, which could be likely improved by trying different settings, such as increasing the number of iterations and fine-tuning the values for prior parameters $\alpha$ and $\beta$. 
    However, note that setting $\alpha$ to 50 divided by the number of topics (\textit{i.e.}, 50/3 for this application), as suggested by \citeauthor{JSSv040i13} (\citeyear{JSSv040i13}), does not increase the accuracy.
\end{itemize}

\section{Conclusion}
\label{Conclusion}

As far as we are aware of, the present work is among the first ones that have addressed the application of text analysis techniques to OpRisk event descriptions, and is the first one that has provided a structured general workflow for such analyses. 
Furthermore, we have complemented the established frameworks of currently applied statistical methods for quantitative data, hence contributing to the construction of a holistic OpRisk management framework. 
Indeed, our ultimate goal is to provide an analytical and measurement framework that considers OpRisk information in its entirety in order to acquire common language and unified understanding of risk.	

We have applied several statistical approaches and models to analyze and cluster operational risk event descriptions using text analysis techniques, in order to identify the main root-causes of such a risk. 
We have enriched the standard text analysis techniques by considering a semantic adjustment capable of dealing with similar concepts expressed by different words. 
The semantic adjustment can be based on word embedding methods, such as \texttt{word2vec}. 
We have used clustering and topic-modelling techniques (\textit{e.g.}, $k$-means and LDA) to validate and support the clustering performed by the analysts. 
On the other hand, it is meaningful to incorporate the information available from the analysts (like, for instance, the number of clusters to be considered) when adopting statistical methods.			 

We have focused on the UniCredit CoRep data set when applying the described text analysis methods and several clustering methods, thus providing a useful comparison that highlights their advantages and limitations.
Our results have allowed to identify two homogeneous clusters of events within the event type “Clients, Products \& Business Practices” concerning “disputes related to irregularities in the interest rates calculation", “disputes related to mortgages in foreign currency", and a residual cluster containing other events within the same event type.
Such results have been validated by statistical indices. 
Notably, they agree with the judgments and the knowledge of skilled analysts in the field.			 
The $k$-means clustering method has provided the highest accuracy to the clusters identified by the analysts. 
However, further analysis of more extended data sets should be performed before drawing conclusions on the best methodologies for these purposes. 					

The proposed framework constitutes a starting point for analyzing operational risk event descriptions. 
It could be improved and extended by focusing on several aspects:
\begin{itemize}
    \item Including the procedure, described in Section \ref{Cluster}, in an analytical loop. 
    At each iteration, event descriptions belonging to the identified clusters can be labeled and then removed from the data set.
    A tag deduction activity can be performed to infer tags of new events from tagged events with similar descriptions by using, \textit{e.g.}, a $k$-nearest neighbors approach.
    \item Systematically applying clustering and topic-modelling techniques to partially automate the identification of the clusters on large data sets.
    \item Employing techniques to drive the selection of the number of relevant clusters or topics (\textit{e.g.}, identifying the number of clusters that maximizes the average silhouette index).
    \item Adopting multidimensional projection techniques, such as Least Square Projection (LSP) (\citeauthor{4378370}, \citeyear{4378370}), or self-organizing maps (SOM) (\citeauthor{Pacella_2016}, \citeyear{Pacella_2016}), to preserve neighborhood relations and improve cluster identification.
    \item Adopting other word embedding techniques, such as GloVe (\citeauthor{Pennington_2014}, \citeyear{Pennington_2014}) or BERT (\citeauthor{Kaliyar_2020}, \citeyear{Kaliyar_2020}) and training them on large operational risk data sets. 
\end{itemize}

%
%


\section*{Acknowledgements}

We are grateful to two anonymous reviewers and the editor in charge of our manuscript for constructive criticism, suggestions, and additional insights.

We thank Ruben Binda and Ennio Menicucci (Group Non-Financial Risks, UniCredit S.p.A.) for their useful hints and continuous support.

We also thank Roberto Boselli, Lorenzo Malandri and Fabio Mercorio (University of Milano Bicocca), Maurizio Romano and Gianpaolo Zammarchi (University of Cagliari) for the valuable discussions on the topics argued in this paper.


\section*{Declaration of interests}

The authors report that there are no competing interests to declare.

\renewcommand*{\bibfont}{\small}

\small{ 
\printbibliography 
}

\end{document}